\documentclass[12pt]{iopart}
\usepackage{iopams}
  
\usepackage{graphicx}
\usepackage{epstopdf}

\usepackage[mediumspace,mediumqspace,squaren]{SIunits}

\usepackage{subcaption}
\usepackage{subfig}

\begin{document}

\title[FIDA observations of fishbones in MAST]{Fast-ion Deuterium Alpha spectroscopic observations of the effects of fishbones in the Mega-Ampere Spherical Tokamak}

\author{O M Jones$^{1,2}$, C A Michael$^2$\footnote{Present address: Plasma Research Laboratory, Research School of Physical Science and Engineering, Australian National University, Canberra, ACT 0200, Australia}, K G McClements$^2$, N J Conway$^2$, B Crowley$^2$, R J Akers$^2$, R J Lake$^{3,2}$, S D Pinches$^2$\footnote{Present address: ITER Organization, Route de Vinon sur Verdon, 13115 St. Paul-lez-Durance, France} and the MAST team$^2$}

\address{$^1$Department of Physics, Durham University, South Road, Durham DH1 3LE, UK}
\address{$^2$EURATOM/CCFE Fusion Association, Culham Science Centre, Abingdon, Oxon OX14 3DB, UK}
\address{$^3$Centre for Fusion, Space and Astrophysics, Department of Physics, University of Warwick, Coventry, W. Midlands CV4 7AL, UK}

\ead{owen.jones@ccfe.ac.uk}

\begin{abstract}
Using the recently-installed Fast-Ion Deuterium Alpha (FIDA) spectrometer, the effects of low-frequency $(20-50\usk\kilo\hertz)$ chirping energetic particle modes with toroidal mode number $n\geq1$ on the NBI-driven fast-ion population in MAST plasmas are considered. Results from the FIDA diagnostic are presented and discussed in the light of the present theoretical understanding of these modes, known as fishbones, in plasmas with reversed shear. Measurements of the fast-ion population reveal strong redistribution of fast ions in both real and velocity space as a result of the fishbones. Time-resolved measurements throughout the evolution of a fishbone show radial redistribution of fast ions with energies up to 95\% of the primary beam injection energy. Correlations between changes in the FIDA signal and the peak time derivative of the magnetic field perturbation are observed in a limited range of operating scenarios. The transient reduction in signal caused by a fishbone may in some cases reach 50\% of the signal intensity before mode onset.
\end{abstract}

\pacs{52.35.Py, 52.50.Gj, 52.55.Fa, 52.55.Pi, 52.55.Tn, 52.70.Kz}



\section{Introduction}\label{sec:intro}
Confinement of fast ions is of crucial importance for the performance of future burning plasma reactors. As a next step device, ITER will derive substantial heating from energetic alpha particles generated by fusion reactions, but will also be reliant on auxiliary heating including neutral beam injection (NBI) and ion cyclotron resonance heating (ICRH) \cite{Kritz2011}. Present devices, in which fusion-born fast ions generally provide only a small contribution to the overall power balance, nonetheless provide test-beds on which techniques to improve confinement of energetic particles produced both by fusion reactions and by auxiliary heating systems may be proven \cite{Duong1993, Turnyanskiy2009}. Here we consider results from the Mega-Ampere Spherical Tokamak (MAST). MAST is a mid-size, low aspect ratio tokamak with major radius $R\approx0.9\usk\metre$ and minor radius $a\approx0.6\usk\metre$. A typical pulse lasts $\lesssim0.5\usk\second$, and has a flat-top plasma current of $400\usk\kilo\ampere - 900\usk\kilo\ampere$ and a toroidal field on axis of $0.40\usk\tesla - 0.55\usk\tesla$. The bulk plasma species is deuterium. Core electron densities and temperatures are of order $10^{19}\usk\metre^{-3}$ and $1\usk\kilo\electronvolt$. Fast ions in MAST are generated by two deuterium neutral beam injection (NBI) systems, each of which is capable of injecting up to $2.5\usk\mega\watt$ of NBI power at energies of $60\usk\kilo\electronvolt - 75\usk\kilo\electronvolt$. These beams inject in an anti-clockwise direction when viewed from above, in the direction of the plasma current and toroidal rotation under normal operating scenarios. The tangency radius of both beams is $R_{\text{tan}}=0.7\usk\metre$.

In recent years, the interaction of fast ions with MHD modes has been the subject of extensive research. On DIII-D, transport of fast ions caused by sawteeth, Alfv\'en eigenmodes and microturbulence has been observed \cite{Pace2011}. Studies on ASDEX Upgrade have found significant losses of fast ions correlated with reversed-shear and toroidicity-induced Alfv\'en eigenmodes (RSAEs and TAEs) \cite{Garcia-Munoz2011}. Drops in the neutron emission from NSTX, which is principally caused by fusion reactions between NBI-produced fast deuterons and thermal deuterons, have been observed when energetic particle modes (EPMs) are active, as well as during periods of large-amplitude, bursting TAE activity \cite{Fredrickson2006}. Note that here we define EPMs to be fast-particle-driven internal kink modes with a dominant poloidal mode number $m=1$, and with fundamental toroidal mode number $n=1$; higher toroidal harmonics may also be present. Fast-ion transport induced by fishbones has recently been reported on JET, where the observed changes in neutron rate were compared with modelling \cite{PerezvonThun2012}. Finally, on MAST, the recently-commissioned collimated neutron camera has been used to detect changes in the radial emissivity profile of neutrons caused by sawteeth and EPMs \cite{Cecconello2012}. A recent topical review by Breizman and Sharapov \cite{Breizman2011} summarises the theoretical progress of the last decade in the study of energetic particles in fusion plasmas, in particular their interaction with Alfv\'en eigenmodes and energetic particle modes.

On MAST, the fast-ion redistribution and losses caused by energetic particle modes are of particular interest for two reasons. Firstly, anomalous (higher than neoclassical) fast-ion redistribution in MAST appears to be due mainly to EPMs, motivating the study of these modes in particular. By contrast, in the case of NSTX, strong fast-ion redistribution due to nonlinear TAE avalanches has also been reported \cite{Fredrickson2006, Podesta2011}. While sawteeth have been observed to cause fast-ion redistribution in ASDEX Upgrade \cite{Geiger2011}, the onset of sawteeth is delayed in MAST discharges with early beam heating and low density \cite{Buttery2004}. This is precisely the set of conditions under which energetic particle modes are driven unstable. Where sawteeth occur, they appear only towards the end of the discharge when the current has penetrated sufficiently to cause the growth of the $q=1$ surface. The onset of sawteeth may furthermore be suppressed by the presence of the $[n,m]=[1,1]$ long-lived mode, which has been postulated to prevent the current penetration required to trigger a sawtooth crash \cite{Chapman2010}. Secondly, in common with NSTX \cite{Fredrickson2006} as well as with conventional large aspect ratio tokamaks \cite{Heidbrink2011a}, chirping EPMs in MAST are observed to set in well before $q_{\min}$ drops to unity and to continue as the $q$-profile evolves, eventually merging into the long-lived mode. This poses challenges for fast-ion confinement in future spherical tokamaks under advanced operating scenarios, where $q$-profiles with $q_{\min}$ significantly above unity and with reversed-shear are expected \cite{Chapman2010}.

\begin{figure}[p]
\begin{centering}
\includegraphics[width=0.85\textwidth]{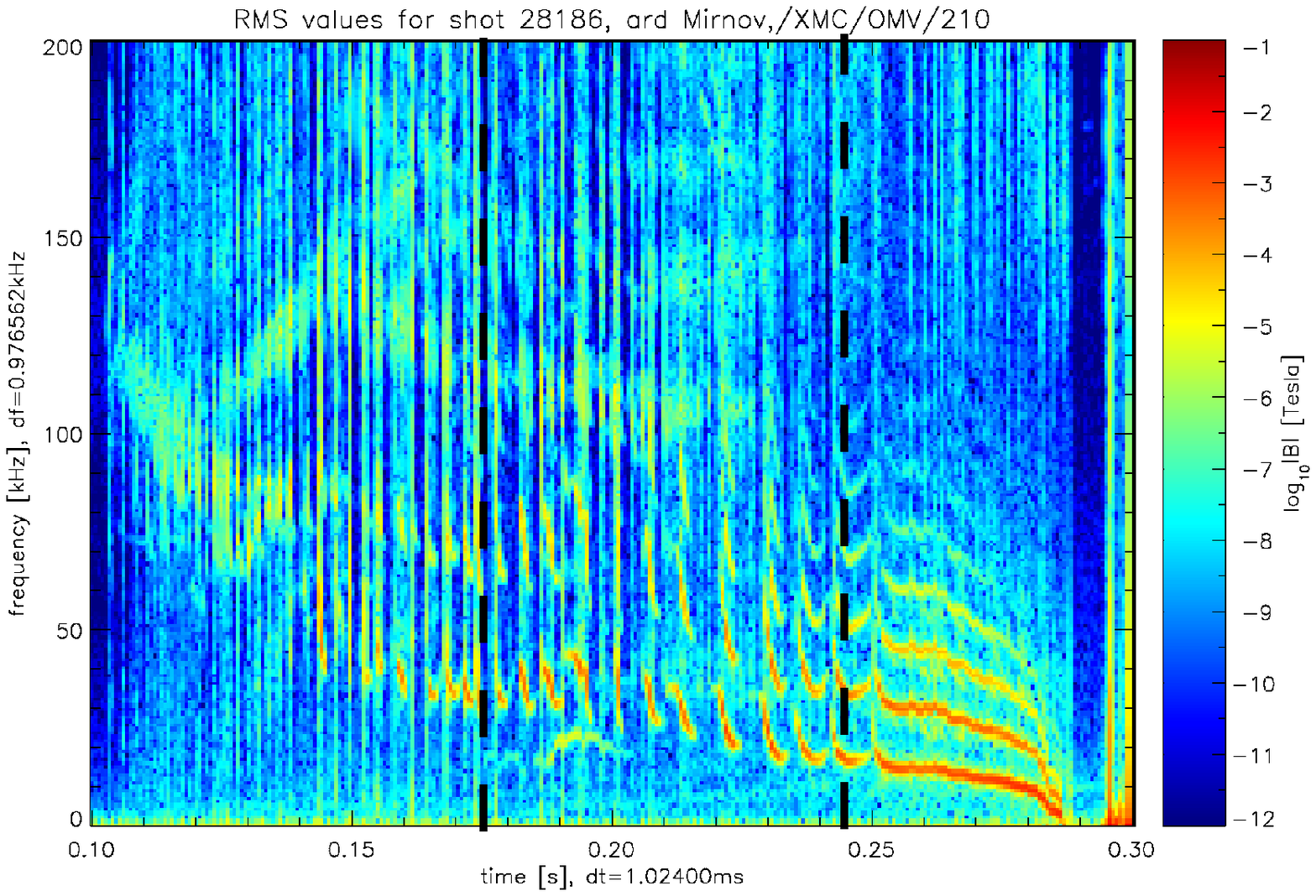}
\caption{Typical magnetic spectrogram of a beam-heated MAST plasma, from a Mirnov coil on the outboard midplane. Beam switch-on occurs at $0.10\usk\second$. Initial bursting activity at TAE frequencies ($\sim100\usk\kilo\hertz$), and chirping modes derived from this activity, last until around $0.15\usk\second$. After this, repeated chirping modes with frequencies below $50\usk\kilo\hertz$ (i.e. fishbones) occur, eventually leading at $0.25\usk\second$ into the long-lived mode. The LLM traces the plasma rotation, gradually reducing in frequency, and disappears shortly before $0.29\usk\second$. Note that higher harmonics of the fishbones and LLM, up to $n=6$, are also observed. The plasma disrupts at $0.30\usk\second$. Dashed lines at $0.175\usk\second$ and $0.245\usk\second$ show the time points chosen for the $q$-profile plots in Figure \ref{fig:qprof}.}
\label{fig:spectrogram}
\end{centering}
\end{figure}

\begin{figure}[p]
\begin{centering}
\includegraphics[width=0.6\textwidth]{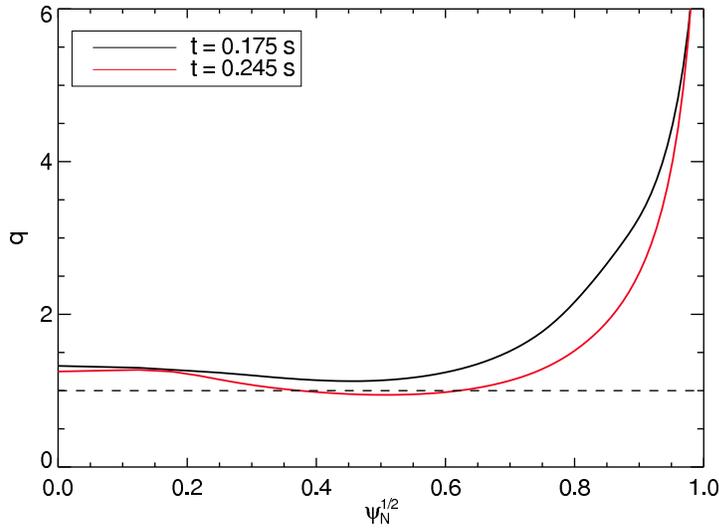}
\caption{Profiles of safety factor $q$ as a function of square root of normalized poloidal flux for MAST shot \#28186 at the times indicated in Figure \ref{fig:spectrogram}. Note that although the profile at $0.245\usk\second$ suggests the existence of a region in which $q$ drops below 1, the absence of sawteeth in the spectrogram shown in Figure \ref{fig:spectrogram} suggests that $q$ may in fact remain marginally above unity. The uncertainty in the MSE-constrained EFIT reconstruction of the $q$-profile is in any case larger than the amount by which $q$ drops below unity.}
\label{fig:qprof}
\end{centering}
\end{figure}

\begin{figure}[p]
\begin{centering}
\includegraphics[width=0.6\textwidth]{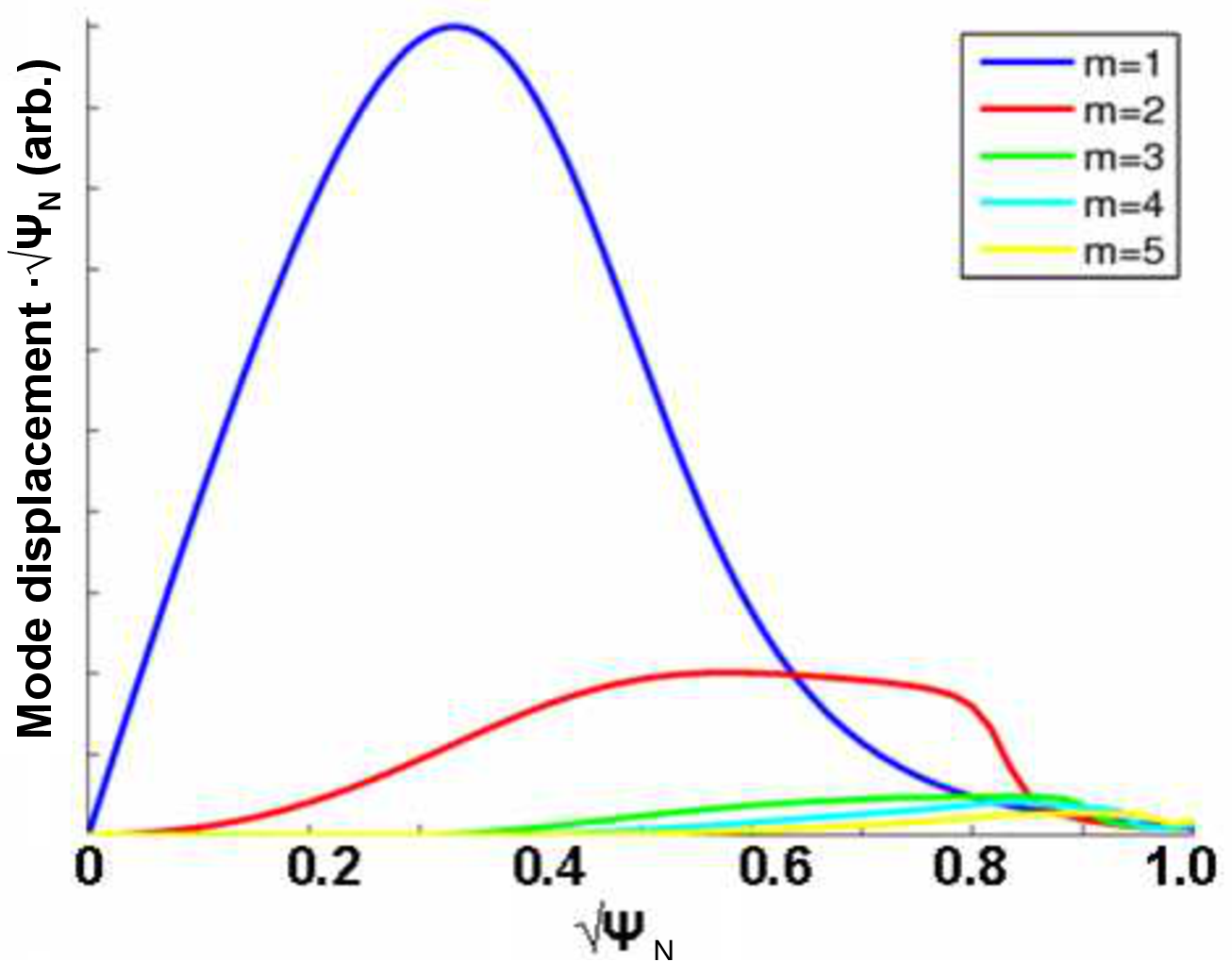}
\caption{Structure of the infernal kink-ballooning mode in a typical low-shear MAST equilibrium, computed with the ideal MHD stability code MISHKA with scalar fast-ion pressure included. The figure illustrates the displacement of poloidal harmonics of the $n=1$ mode as a function of the square root of normalized poloidal flux.}
\label{fig:structure}
\end{centering}
\end{figure}

In MAST, shortly after the disappearance of chirping TAEs, energetic particle modes similar to the $[n,m]=[1,1]$ fishbones first identified in PDX \cite{McGuire1983} are observed to chirp downward in frequency, typically from 40 to 20\usk\kilo\hertz. The fishbones originally observed in PDX were associated with sawtooth-like phenomena in soft X-ray emission \cite{McGuire1983}, implying the presence of a $q=1$ surface inside the plasma. In contrast, the fishbones observed in MAST occur even in the absence of a $q=1$ surface \cite{Pinches2012}. Nonetheless, the similarities between the two cases in terms of the associated magnetic coil signals and mode frequency evolution are sufficient that these modes are hereafter termed `fishbones'. These fishbones in many cases later evolve into the long-lived mode (LLM), which is believed to be an ideal internal kink mode \cite{Chapman2010}. The LLM frequency evolves slowly in time compared to the fishbone frequency, tracking the plasma rotation. Figure \ref{fig:spectrogram} shows the temporal evolution and frequencies typical of each of these modes, while Figure \ref{fig:qprof} shows the reversed-shear $q$-profiles associated with fishbones at two different times during the same shot. These $q$-profiles are derived from an EFIT equilibrium reconstruction constrained by data from the Motional Stark Effect (MSE) diagnostic. The mode structure of a typical fishbone (not necessarily from shot \#28186, but from a double-null diverted MAST shot with a similar low-shear $q$-profile) calculated with the ideal MHD stability code MISHKA is shown in Figure \ref{fig:structure}. The difference between this mode and the original PDX fishbones, which displayed a `top-hat' structure with the $m=1$ component localised close to the $q=1$ surface \cite{McGuire1983}, is evident. The MAST fishbones are identified with the $n=1$ infernal kink-ballooning mode \cite{Pinches2012, Kolesnichenko2006}, which has a dominant $m=1$ harmonic residing in the broad, low-shear region in the core, and harmonics with higher poloidal mode numbers residing at larger radii.

The present work draws upon data obtained using the recently-commissioned Fast-Ion Deuterium Alpha (FIDA) spectrometer on MAST. The effect of fishbones on the fast-ion population in MAST plasmas has been studied using the radially-scanning collimated neutron detector \cite{Cecconello2012}, but the unique ability of the FIDA diagnostic to resolve details of the fast-ion distribution in velocity space as well as real space significantly extends this study. It is not our intention to present a detailed description of the diagnostic, since such a description will be provided in a separate paper, currently in preparation \cite{Michael2013}. Readers interested in the background of the experimental method are referred to overviews of similar installations on DIII-D \cite{Heidbrink2010} and NSTX \cite{Bortolon2010}. Only a brief outline will be presented here. It is worth noting that while the FIDA installation on ASDEX Upgrade has been used to observe changes in the fast-ion distribution associated with sawtooth crash events \cite{Geiger2011}, the effects of fishbones have not yet been studied in detail with a FIDA diagnostic.

\section{FIDA spectroscopy on MAST}\label{sec:method}
In common with FIDA diagnostics on NSTX \cite{Bortolon2010}, DIII-D \cite{Muscatello2010}, ASDEX Upgrade \cite{Geiger2011}, TEXTOR \cite{Delabie2008} and LHD \cite{Osakabe2008}, the FIDA spectrometer installed on MAST relies on charge exchange of fast ions with beam neutrals to generate its signal. The measurement is nominally localised to the point of intersection between the line of sight and the neutral beam, but the spatial extent of the beam footprint and surrounding `halo' neutrals, produced by charge exchange between injected beam neutrals and thermal deuterons, limits the spatial resolution along the line of sight to approximately $20\usk\centi\metre$. Perpendicular spatial resolution is set by the velocity of the reneutralized fast ion and the decay time of the $n=3\rightarrow2$ transition, and is approximately $2\usk\centi\metre$. On MAST, where the fast-ion density is typically a significant fraction of the plasma density and the relative energies of fast ions and beam neutrals are favourable to the required charge exchange reaction, a temporal resolution of $0.3\usk\milli\second$ is attainable. The sensitivity of a given viewing chord in velocity space (or, equivalently, in energy and pitch space) is dependent on the angles between the chord, the magnetic field at the point of beam intersection, and the velocity vector of the injected beam neutrals. Each viewing chord is sensitive to reneutralized fast ions in a region bounded by an arc in energy/pitch space. Selecting a particular wavelength in the FIDA spectrum corresponds to selecting a minimum energy $E_{\min}$, as only a deuteron of at least that energy could emit D$_{\alpha}$ light with the chosen redshift after being reneutralized. Deuterons of higher energies also contribute to the signal at this wavelength if they are reneutralized at a specific point on their gyro-orbit. See Reference \cite{Michael2013} for details of these `weight functions' in velocity space.

\begin{figure}[p]
\begin{centering}
\includegraphics[width=0.6\textwidth]{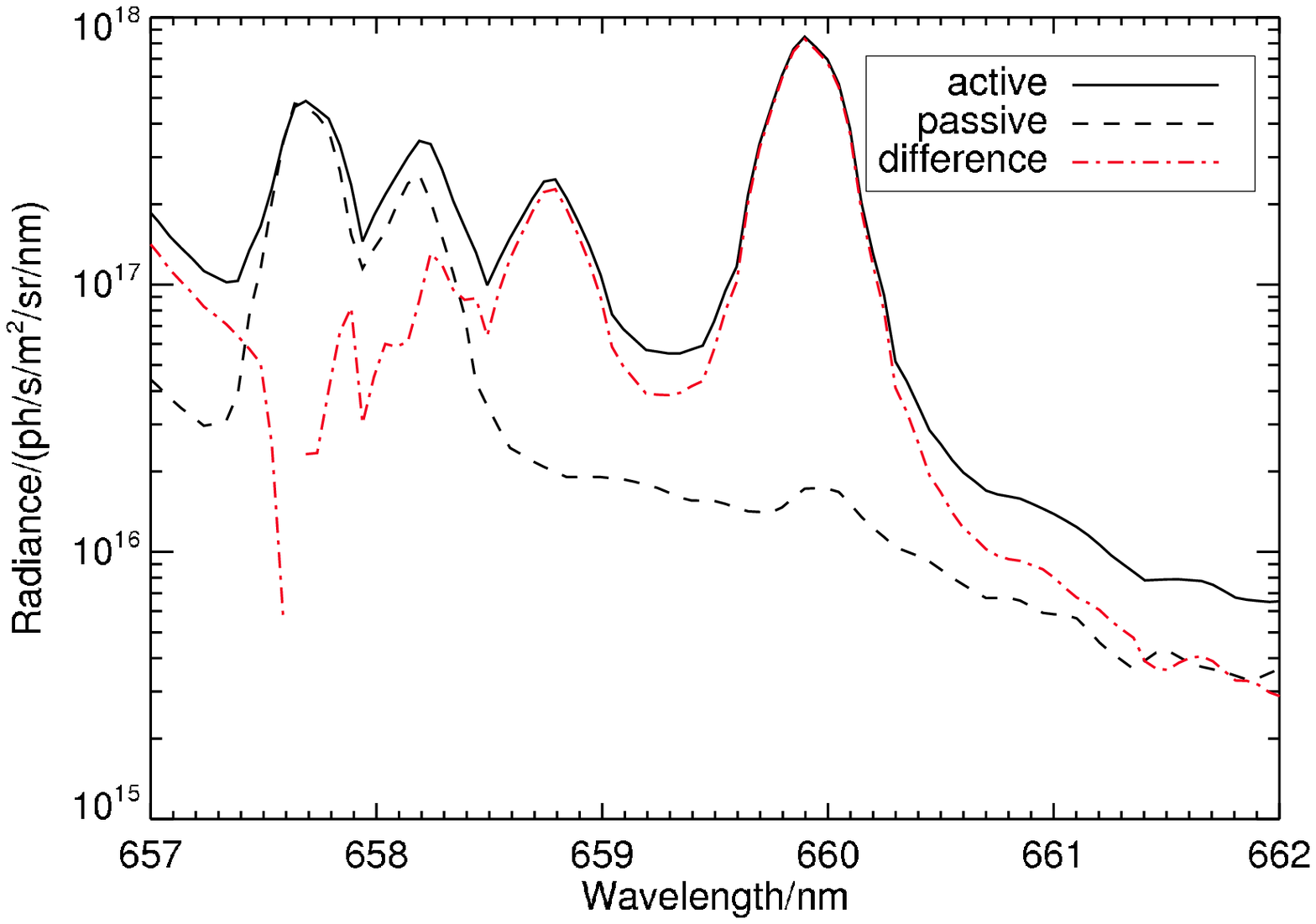}
\caption{Absolutely calibrated data from the FIDA diagnostic for shot \#26970 at $0.180\usk\second$. The active (viewing the SW neutral beam) and passive (looking below the neutral beam) spectra are shown, as is the net spectrum. The carbon lines are present in both active and passive spectra, whereas the full and half-energy beam emission peaks are identified by their absence from the passive spectrum. The third-energy beam emission peak coincides with one of the carbon lines, so background subtraction is not completely reliable, but a peak in the net signal at this wavelength is nonetheless observed. Note the offset from zero of the net spectrum even at $662.0\usk\nano\metre$, where there is no FIDA signal or impurity line radiation; this is caused by the active and passive lines of sight viewing different volumes of plasma, and so observing different amounts of bremsstrahlung.}
\label{fig:spectra}
\end{centering}
\end{figure}

A transmission grating spectrometer is employed on MAST to analyse the spectrum in the region of the Balmer alpha line of deuterium, centred on $656.1\usk\nano\metre$. Features in this wavelength range include the light from `cold' deuterium at the plasma edge, beam emission peaks at the full, half and one-third injection energies, carbon and oxygen impurity line radiation, a background of bremsstrahlung, and a weak FIDA spectrum. These features may be seen in Figure \ref{fig:spectra}. Subtraction of bremsstrahlung and passive FIDA emission arising from charge-exchange of fast ions on edge neutrals is provided to first order by passive views which do not view a neutral beam, but have similar lines of sight to the active (beam-viewing) lenses. Unfortunately the vertical and toroidal displacement of the passive views relative to the active means that bremsstrahlung and passive FIDA are not correctly subtracted in all cases. In the case of bremsstrahlung, which is observed as a flat elevation of the baseline over the spectral region of interest, background subtraction is further enforced by subtracting the offset from zero of the net signal (difference between active and passive) at a selected wavelength. This wavelength is chosen such that there is no impurity radiation and no FIDA signal either, since the fast-ion energy required to Doppler-shift the D$_{\alpha}$ emission to this wavelength is significantly greater than the primary NBI injection energy. Passive FIDA cannot be treated in this way as it has a particular spectral shape depending on the fast-ion distribution at the plasma edge. This leads to systematic errors in background subtraction. The systematic errors particularly affect the signal from the outer viewing chords, with midplane intersection radii from approximately $R=1.25\usk\metre$ outward, where the ratio of edge neutrals to beam neutrals is high and the fast-ion density (and hence magnitude of the signal) is low. Where these errors affect the results presented here, they are acknowledged.

The spectrum shown in Figure \ref{fig:spectra} is obtained using the horizontally-viewing FIDA lens. This lens views co-injected neutrals and co-going fast ions from behind, so the FIDA signal is redshifted. This viewing geometry however also results in large parts of the FIDA spectrum being contaminated by beam emission. A vertical view which sees blueshifted light from reneutralized co-going ions is also available, and is not subject to such contamination, but the signal from this view is too weak to allow identification of transient drops caused by MHD activity. For this reason, only data from the horizontal system are presented here. Uncertainties are derived from a combination of shot noise and an assumed systematic error of $\pm20\%$ in the magnitude of the passive signal, as outlined in Reference \cite{Michael2013}. Shot noise errors in the net signal are typically $5\%$ at the lowest usable FIDA wavelength of $659.5\usk\nano\metre$, where the signal is strongest, and $25\%$ at the highest usable wavelength of $661.5\usk\nano\metre$, where the signal is weakest. At lower wavelengths, the systematic error bar ($20\%$ of the passive signal) dominates since the magnitude of the passive signal is larger at lower wavelengths.

\section{Results and Discussion}\label{sec:results}
\begin{figure}[p]
\begin{centering}
\includegraphics[width=0.6\textwidth]{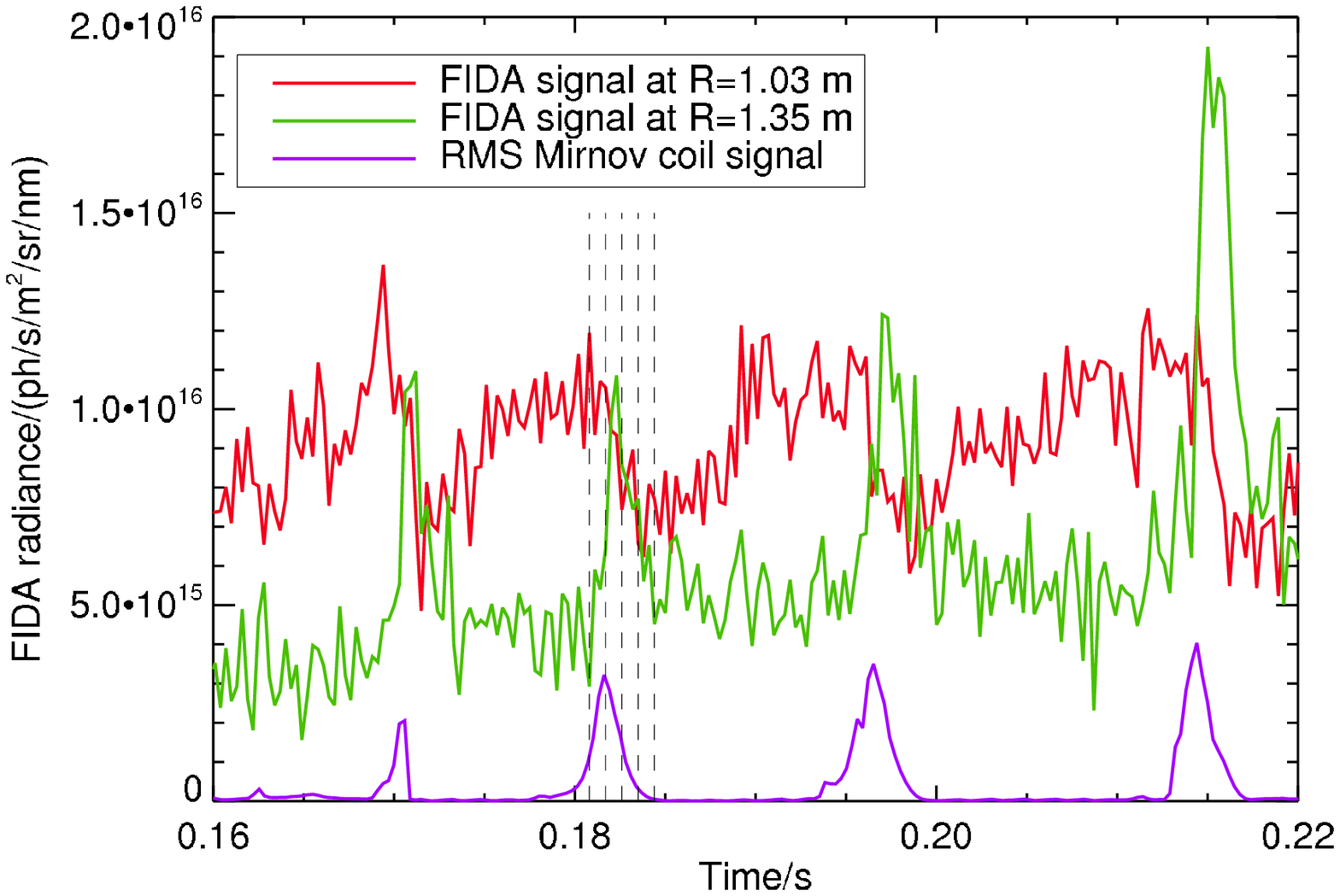}
\caption{Time trace of FIDA data from MAST shot \#26789. The signals from chords which intersect the neutral beam at midplane radii of $1.03\usk\metre$ and $1.35\usk\metre$ are shown, as is the RMS amplitude of the outboard midplane Mirnov coil signal binned over the same time resolution as the FIDA data. Vertical dashed lines indicate the time points chosen for the plots of spectra shown in Figure \ref{fig:spectra26789}. Data are for $\lambda=660.5\usk\nano\metre$.}
\label{fig:trace26789}
\end{centering}
\end{figure}

As seen in Figure \ref{fig:spectrogram}, many MAST discharges show periods of repeated chirping EPMs unaccompanied by other MHD modes. Such periods allow the effects of fishbones on the fast-ion population to be isolated, as is required for the present study. Figure \ref{fig:trace26789} shows a time trace of FIDA data at two different radii during a period with repeated fishbones. Note the drops in the core signal ($R=1.03\usk\metre$) associated with each fishbone burst, followed by a steady recovery between each burst, while the edge signal ($R=1.35\usk\metre$) exhibits large, transient spikes at the time of each fishbone. It is likely that these transient spikes arise largely from the signal from lost fast ions undergoing charge exchange on edge neutrals; the incorrect background subtraction of this passive FIDA signal mentioned in the previous section accounts for the size of the bursts in edge signal. The fact that the core FIDA signal returns to the same value before each fishbone burst is consistent with the limit-cycle model of fishbones, whereby core fast-ion pressure increases until a chirping mode is driven unstable by the pressure gradient. The mode causes redistribution or loss of fast ions, reducing the fast-ion pressure gradient until the mode is stabilised and the cycle is repeated.

\begin{figure}[p]
	\begin{subfigure}[t]{0.5\textwidth}
	\centering
	\includegraphics[width=\textwidth]{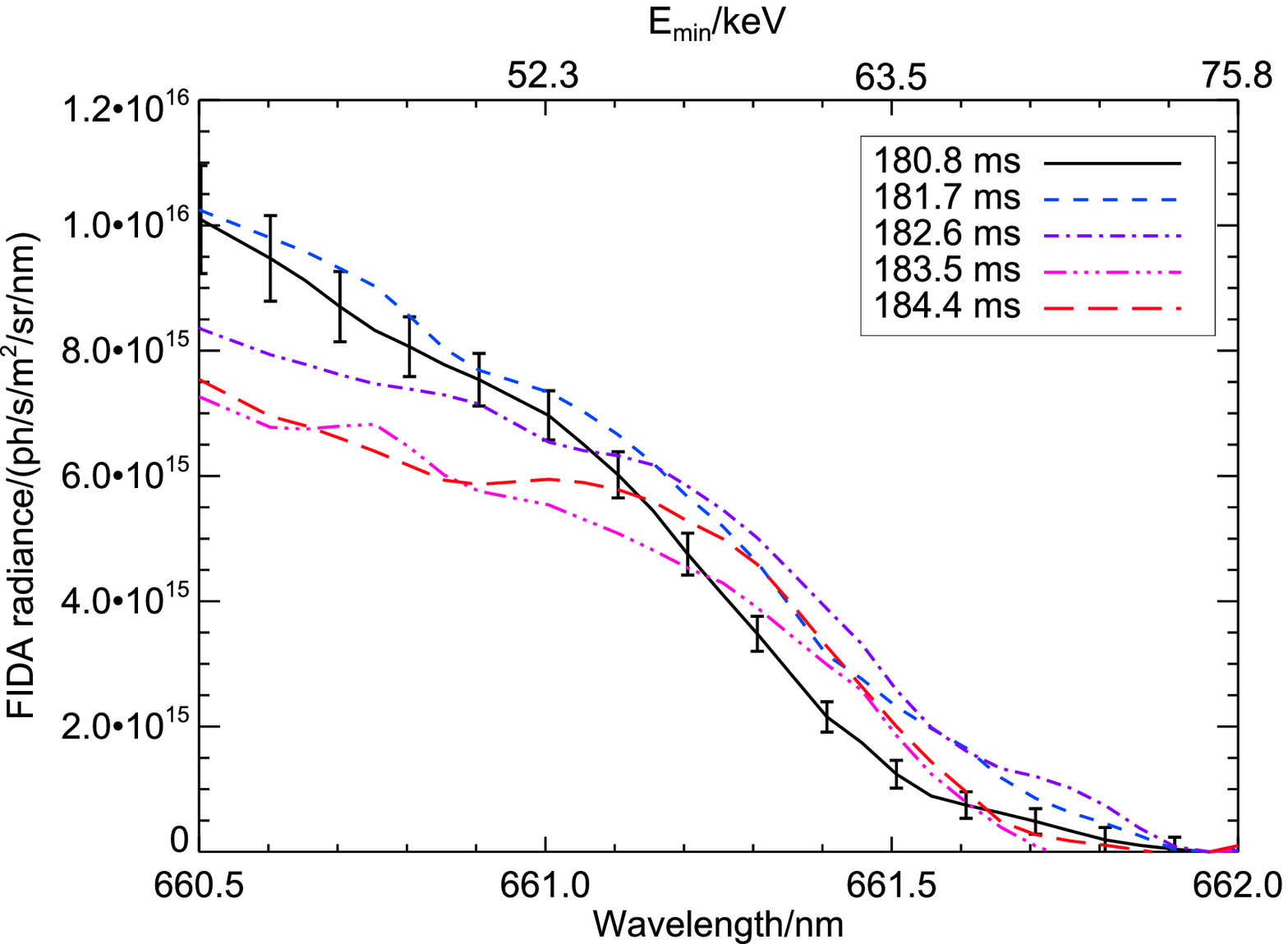}
	\caption{$R=1.03\usk\metre$}
	\label{fig:spectra26789:inner}
	\end{subfigure}
	\begin{subfigure}[t]{0.5\textwidth}
	\centering
	\includegraphics[width=\textwidth]{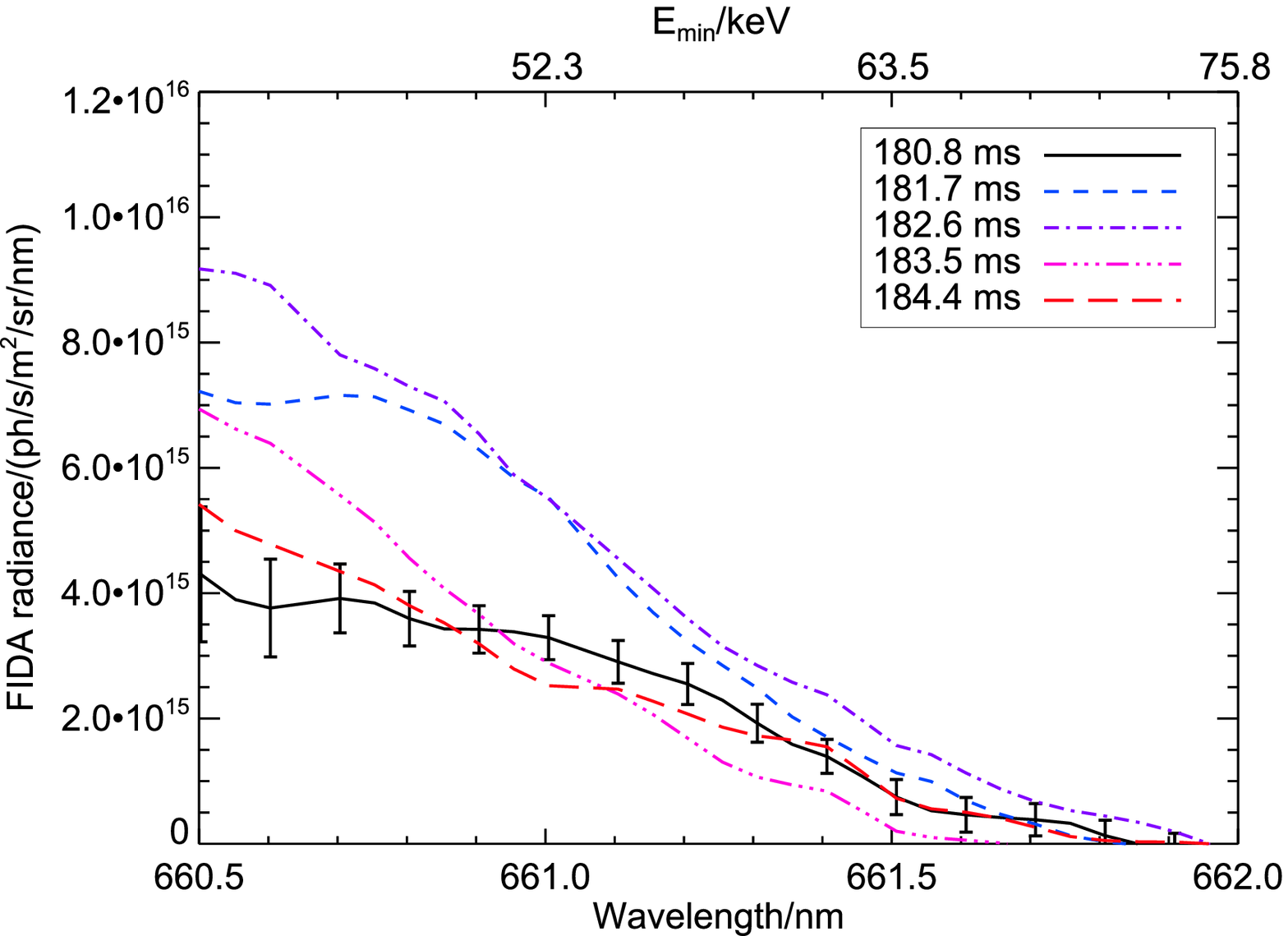}
	\caption{$R=1.35\usk\metre$}
	\label{fig:spectra26789:outer}
	\end{subfigure}
\caption{FIDA spectra from MAST shot \#26789 at the times indicated by dashed lines in Figure \ref{fig:trace26789}, at radii of (a) $1.03\usk\metre$ and (b) $1.35\usk\metre$. For clarity, error bars are shown only for one time slice.}
\label{fig:spectra26789}
\end{figure}

Examining the FIDA spectra around the time of a single burst is instructive. Figure \ref{fig:spectra26789:inner} shows the spectrum of the core channel at $R=1.03\usk\metre$; for a typical MAST plasma with the last closed flux surface located at $R=0.23\usk\metre$ on the inboard and $R=1.41\usk\metre$ on the outboard midplane, and the magnetic axis at $R=0.92\usk\metre$, this radius corresponds to $r/a\approx0.2$. The minimum deuteron energy $E_{\min}$ contributing to the signal at a given wavelength is shown on the top axis of each plot in Figure \ref{fig:spectra26789}. Note that the two beams in this case operate at $56\usk\kilo\electronvolt$ and $64\usk\kilo\electronvolt$. After the time derivative of the magnetic perturbation reaches a maximum (which occurs at $181.7\usk\milli\second$), the signal at wavelengths below $661.1\usk\nano\metre$ drops rapidly ($182.6\usk\milli\second$); the region over which the signal is depleted spreads to encompass all wavelengths up to $661.4\usk\nano\metre$ ($183.5\usk\milli\second$); and finally the signal starts to recover at higher wavelengths, starting at $661.4\usk\nano\metre$ ($184.4\usk\milli\second$). Less may be gleaned from looking at the spectrum at $R=1.35\usk\metre$ ($r/a\approx0.9$) in Figure \ref{fig:spectra26789:outer}, due to the large transient signal caused by incorrect subtraction of passive FIDA, but note that the variation in signal is largely limited still to wavelengths below $661.4\usk\nano\metre$. The corresponding line-of-sight deuteron energy at this wavelength is $61\usk\kilo\electronvolt$.

\begin{figure}[p]
\begin{centering}
\includegraphics[width=0.6\textwidth]{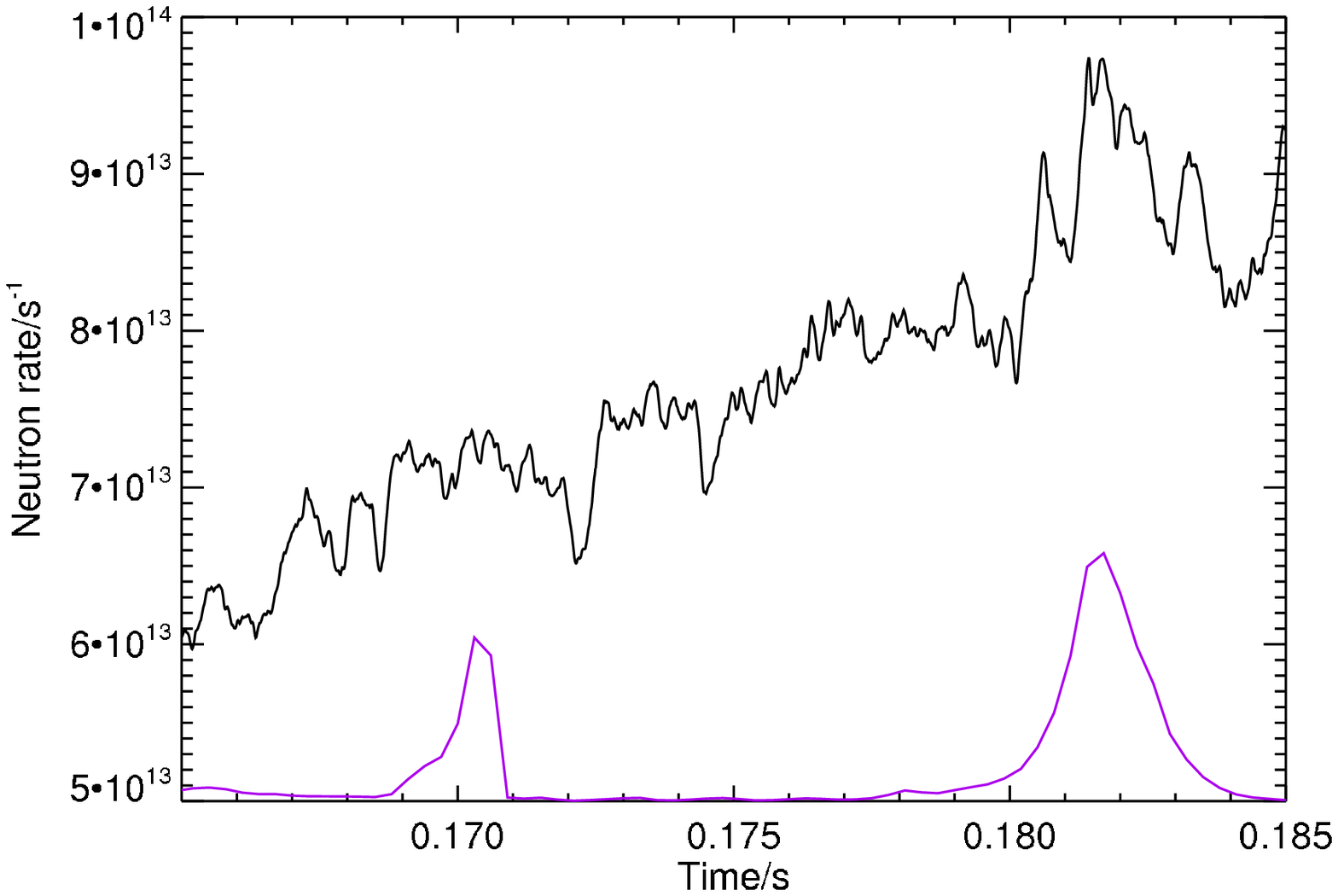}
\caption{Volume-integrated neutron rate from MAST shot \#26789 around the time of the fishbone selected for analysis in Figure \ref{fig:spectra26789}. Note the increase in the neutron rate at the onset of the mode, which implies a growth in the population of fast ions near the injection energy. This is supported by the behaviour of the FIDA signal in Figure \ref{fig:spectra26789}. The RMS amplitude of the outboard midplane Mirnov coil signal is also shown for reference.}
\label{fig:neutrons}
\end{centering}
\end{figure}

The time-resolved behaviour of the core emissivity seen in Figure \ref{fig:trace26789} is consistent with a scenario in which the fishbone displaces a resonant population of fast ions while the neutral beam `pumps' the fast ion population from high energies. At large amplitude, the sink term due to the MHD mode is larger than the source term due to the NBI, so the fast-ion density in a given resonant portion of phase space decreases. With the fast-ion population depleted, the pressure gradient drive is no longer sufficient to sustain the mode, and the fishbone decays in amplitude. Closer inspection of Figure \ref{fig:spectra26789:inner}, however, reveals this interpretation to be problematic. Instead of the expected behaviour of the fishbone redistributing high-energy ions before chirping down in frequency, systematically depleting the population of low-energy ions, the spectrum in Figure \ref{fig:spectra26789:inner} seems to be depleted at lower energies before it is depleted at higher energies. It is interesting to note that the core FIDA signal at wavelengths above $661.1\usk\nano\metre$ increases at the onset of the fishbone. The volume integrated neutron rate, which is generated predominantly by the highest energy fast ions due to the strong energy dependence of fusion cross-section \cite{Bosch1992}, also increases significantly over the same period before dropping as the mode evolves. The neutron rate from the MAST fission chamber preceding and during the selected event is shown in Figure \ref{fig:neutrons}; the neutron rate increases rapidly at the onset of the fishbone, before the perturbation causes fast ions to be expelled from the core and the neutron rate drops again. The cause of this rapid, transient rise in neutron rate is unclear, but its occurrence is consistent with the behaviour observed in the FIDA data at large redshifts. The counterintuitive behaviour of the core FIDA signal in this particular instance motivates future investigation of the variability of FIDA evolution between different events.

\begin{figure}[p]
\begin{centering}
\includegraphics[width=0.6\textwidth]{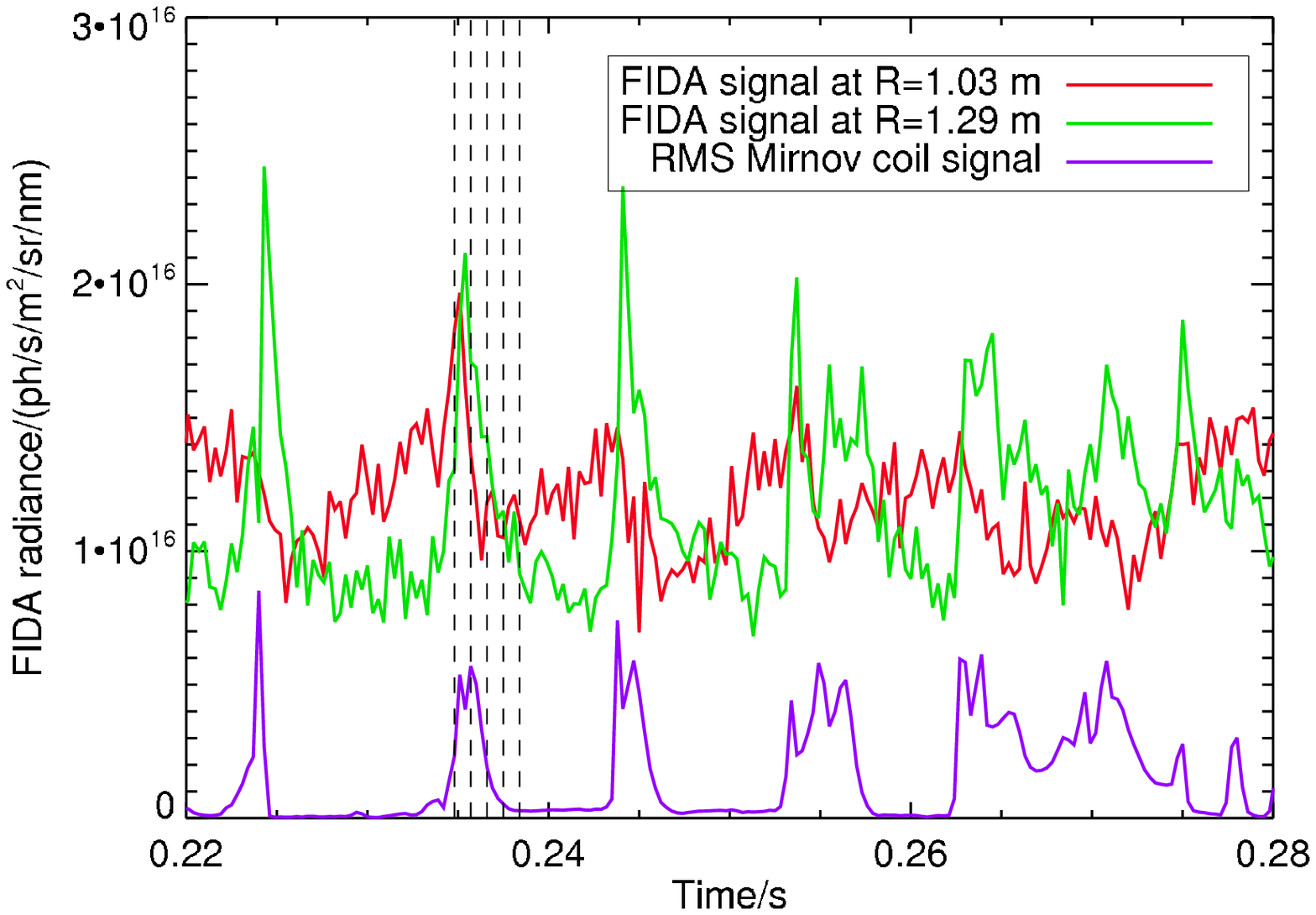}
\caption{Time trace of FIDA data from MAST shot \#26863. The signals from chords which intersect the neutral beam at midplane radii of $1.03\usk\metre$ and $1.29\usk\metre$ are shown, as is the RMS amplitude of the outboard midplane Mirnov coil signal binned over the same time resolution as the FIDA data. Vertical dashed lines indicate the time points chosen for the plots of spectra shown in Figure \ref{fig:spectra26863}. Data are for $\lambda=660.5\usk\nano\metre$.}
\label{fig:trace26863}
\end{centering}
\end{figure}

\begin{figure}[p]
	\begin{subfigure}[t]{0.5\textwidth}
	\centering
	\includegraphics[width=\textwidth]{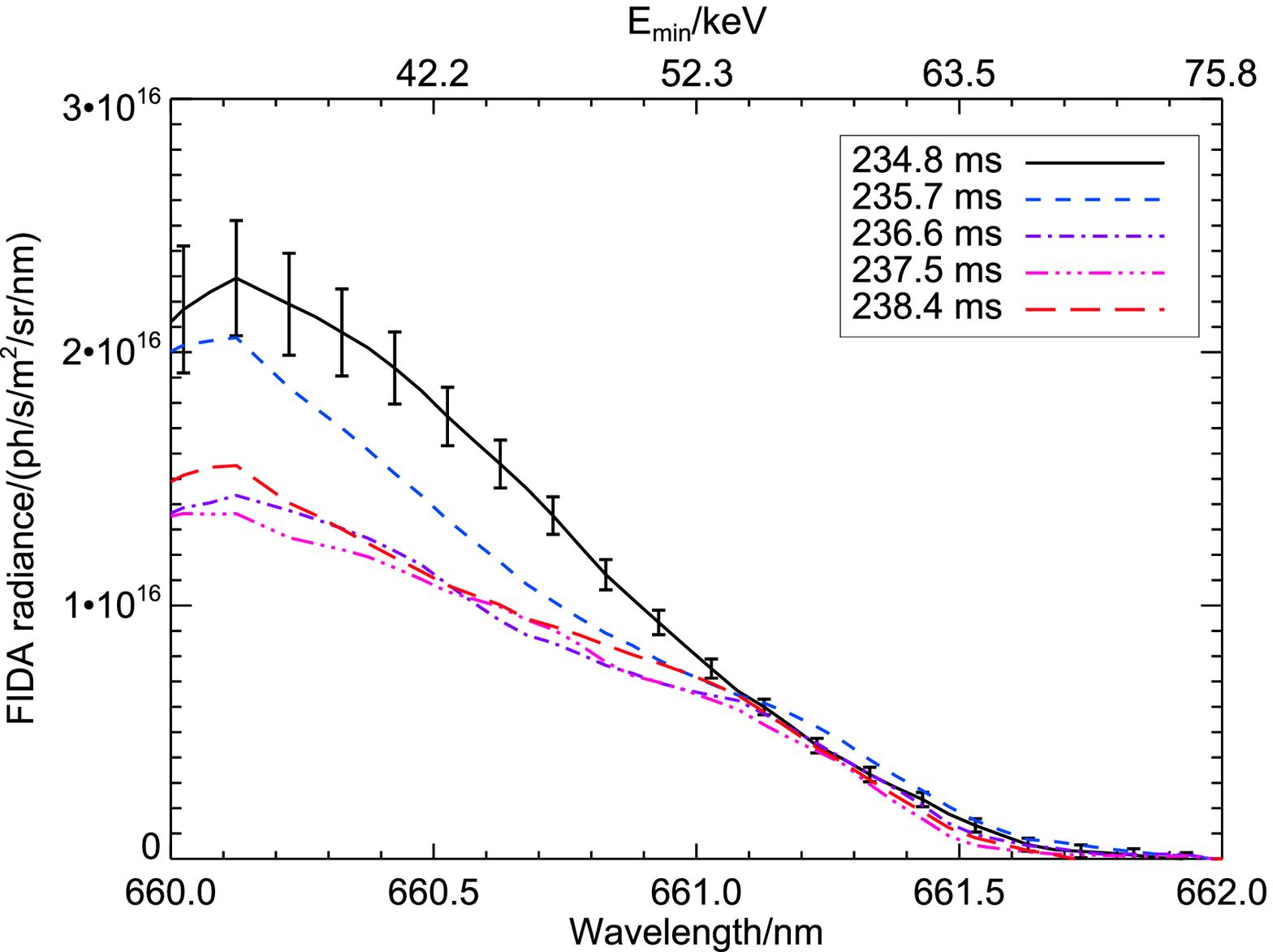}
	\caption{$R=1.03\usk\metre$}
	\label{fig:spectra26863:inner}
	\end{subfigure}
	\begin{subfigure}[t]{0.5\textwidth}
	\centering
	\includegraphics[width=\textwidth]{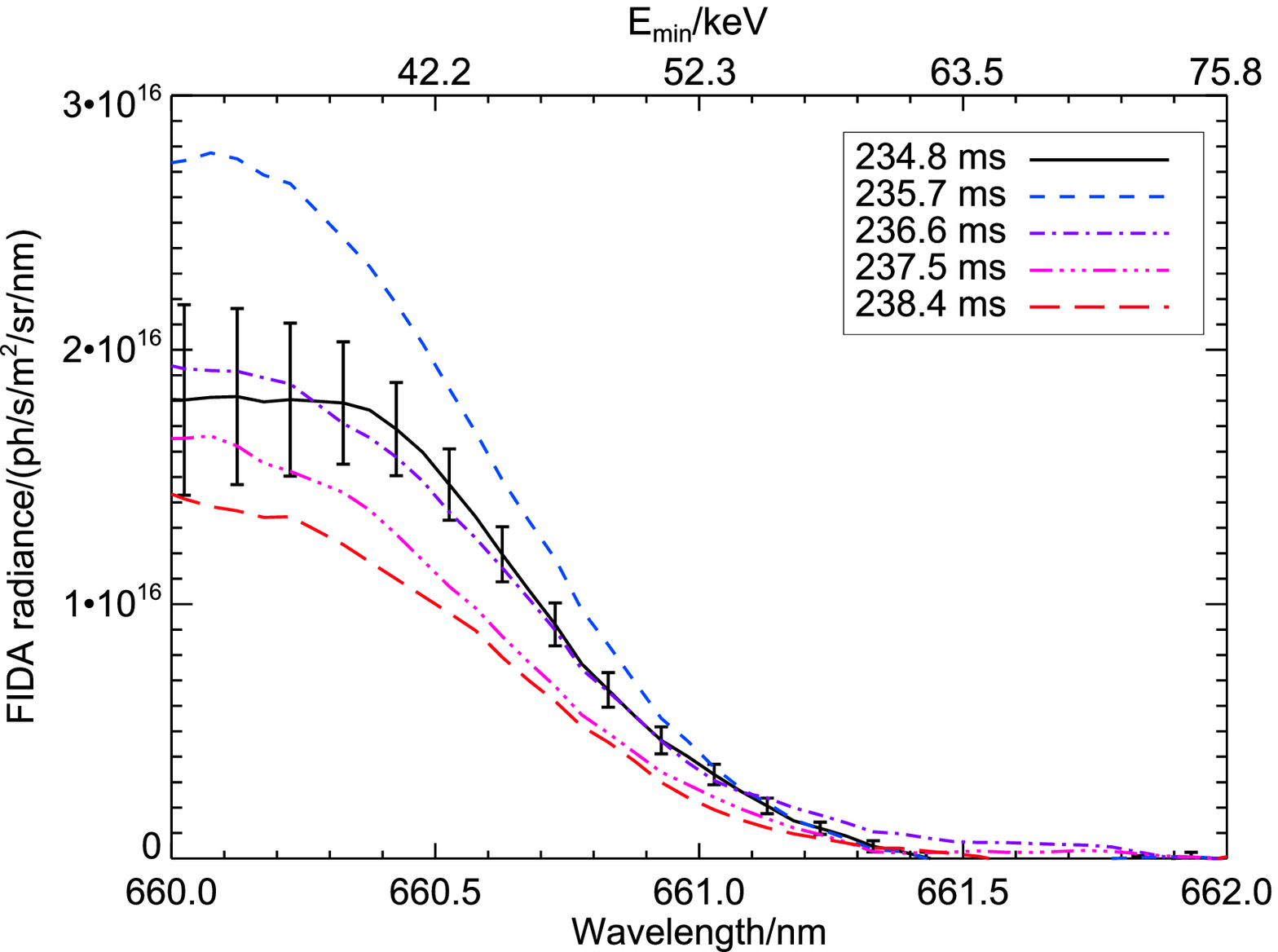}
	\caption{$R=1.29\usk\metre$}
	\label{fig:spectra26863:outer}
	\end{subfigure}
\caption{FIDA spectra from MAST shot \#26863 at the times indicated by dashed lines in Figure \ref{fig:trace26863}, at radii of (a) $1.03\usk\metre$ and (b) $1.29\usk\metre$. The abscissae in these plots extend to lower wavelengths than those in Figure \ref{fig:spectra26789} because in this case a physical mask was used to block beam emission which would otherwise contaminate wavelengths below $660.5\usk\nano\metre$.}
\label{fig:spectra26863}
\end{figure}

Figures \ref{fig:trace26863} and \ref{fig:spectra26863} show the FIDA signal around the time of a fishbone during a different shot. The particular fishbone selected for analysis in this case causes a more rapid drop in core FIDA signal than that in Figure \ref{fig:spectra26789}, but appears to affect only wavelengths up to $661.1\usk\nano\metre$. This may be expected however, as in the case of shot \#26863 both neutral beams inject deuterium with a primary energy of $60\usk\kilo\electronvolt$; since the dominant process determining the dynamics of fast ions at high energies is collisional slowing-down on electrons, a resonant energetic particle mode would only be strongly driven by, and hence only displace, the fast ions with energies below the injection energy. The equivalent lowest energy $E_{\min}$ for D$_\alpha$ at $661.1\usk\nano\metre$ is $54\usk\kilo\electronvolt$. In contrast with the fishbone selected from shot \#26789, no increase in the neutron rate or the high-energy FIDA signal is seen during mode growth in this case. The data from the edge channel are still affected by the transient passive FIDA spikes, although in this case these spikes seem to peak abruptly at the onset of the chirping mode, then decay slowly to the background level before the subsequent burst. Both of the shots selected for analysis here, \#26789 and \#26863, are high-density shots which use MAST's two neutral beams for a total injected power of $3\usk\mega\watt$, although one of the two beams comes on slightly earlier in shot \#26789. Line-integrated density, plasma current and normalized beta are all slightly higher in shot \#26789, which results in the shot entering H-mode at $0.215\usk\second$. Shot \#26863 remains in L-mode for its duration.

According to the kinetic theory of the interaction of particles with chirping modes put forward by Hsu \etal \cite{Hsu1994}, the frequency chirp of the fishbone is responsible for expanding the region of velocity space affected by radial transport. By sweeping in frequency, the mode may redistribute particles in phase space. The position of the particle in phase space governs the strength of its interaction with the wave, which in turn governs the rate of convective radial transport. It remains to be seen whether introducing convective radial transport into existing transport models which reconstruct the fast-ion distribution, such as TRANSP, can account for the redistribution seen in MAST; this will be the subject of future investigation.

\begin{figure}[p]
\begin{subfigure}[t]{0.5\textwidth}
\centering
\includegraphics[width=\textwidth]{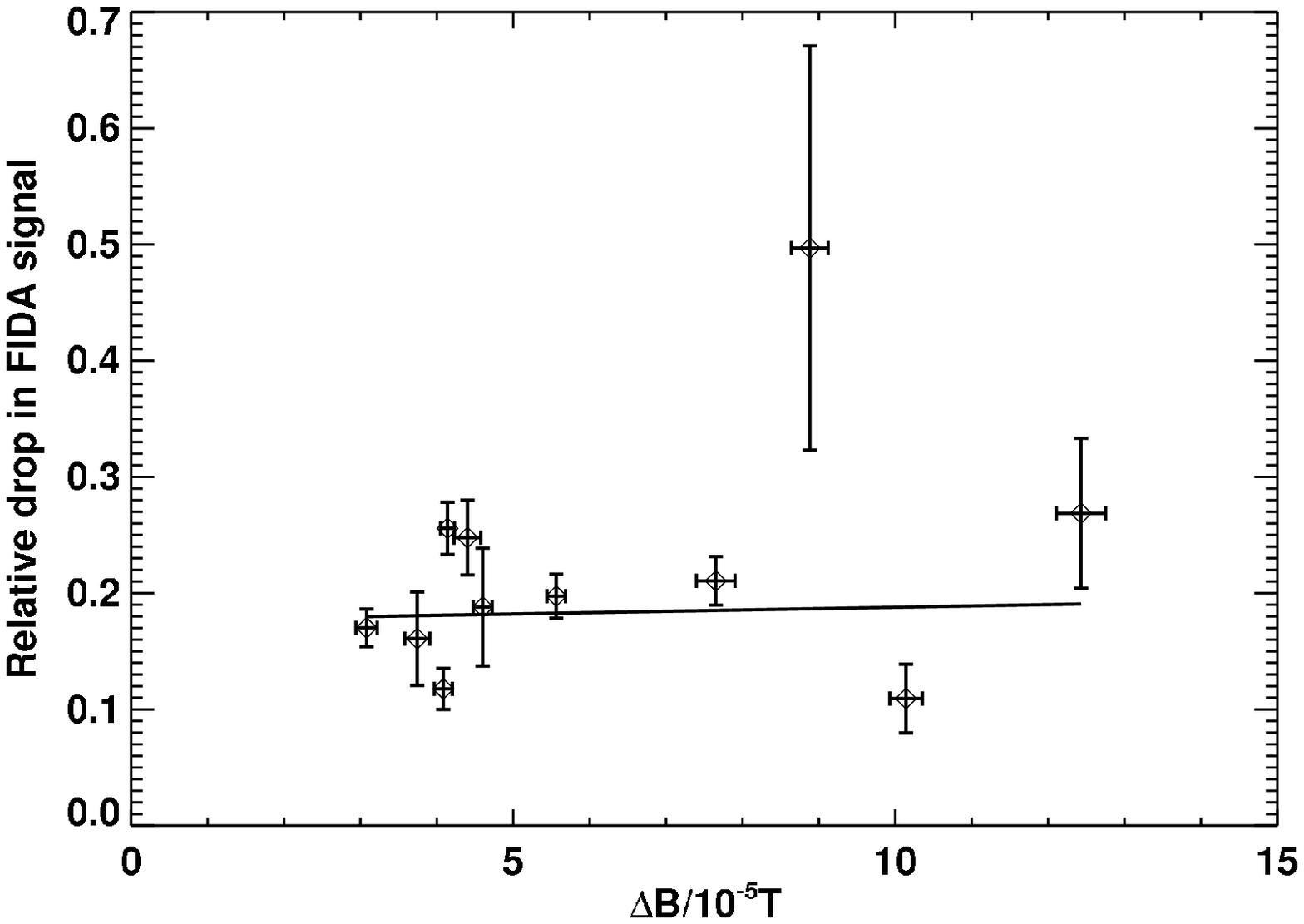}
\caption{FIDA signal drop versus $\Delta B_{\theta}$.}
\label{fig:deltabcorr}
\end{subfigure}
\begin{subfigure}[t]{0.5\textwidth}
\centering
\includegraphics[width=\textwidth]{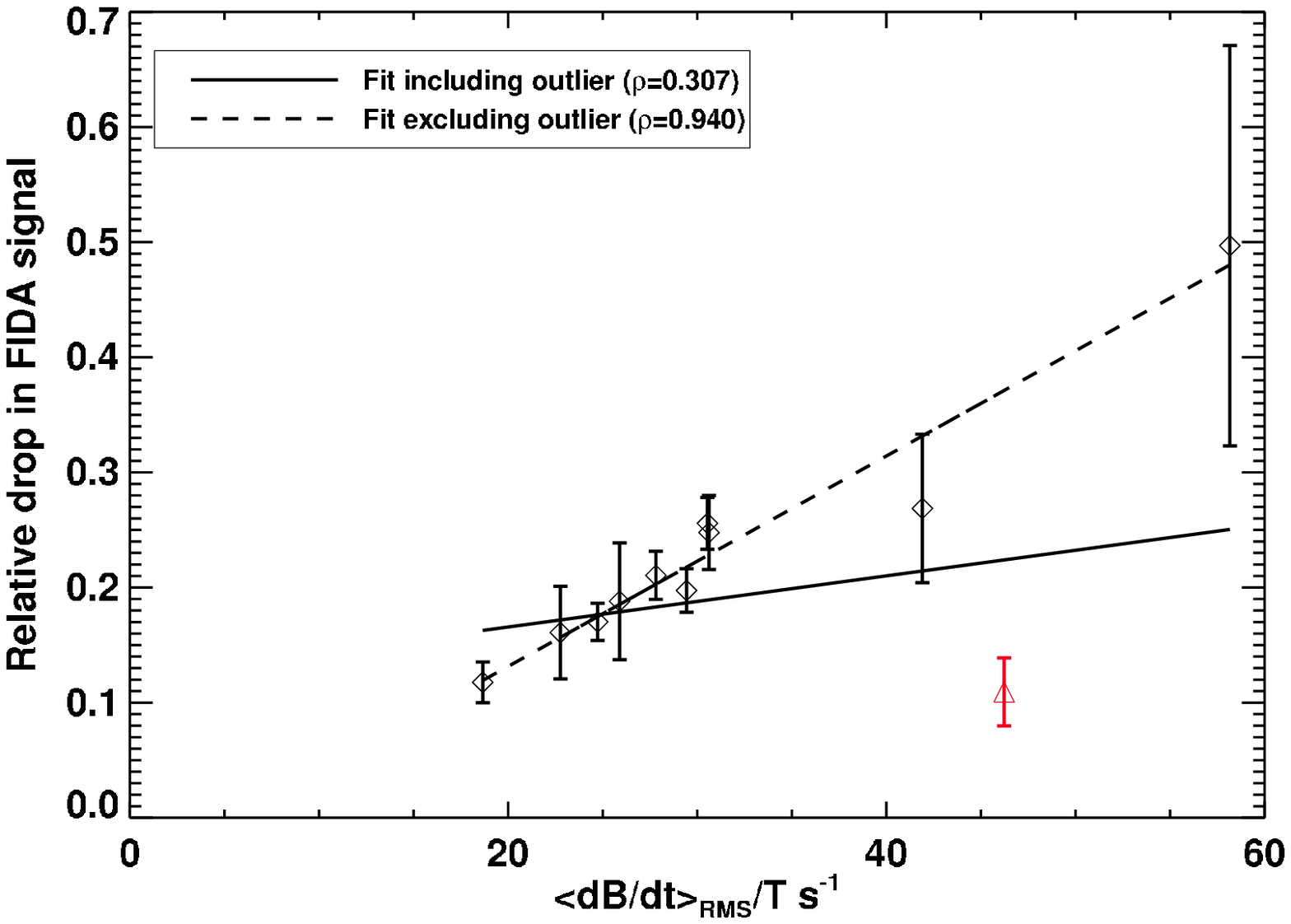}
\caption{FIDA signal drop versus $\langle \partial B_{\theta}/\partial t \rangle_{\text{RMS}}$.}
\label{fig:bdotcorr}
\end{subfigure}
\caption{Correlation plots of relative drop in FIDA signal at $R=1.03\usk\metre$, $\lambda=660.5\usk\nano\metre$, versus magnetic fluctuation amplitude measured by a Mirnov coil on the outboard midplane. Subfigure (a) shows the relative change of FIDA signal against the magnetic perturbation amplitude, $\Delta B_{\theta}$. Subfigure (b) instead takes the maximum amplitude of the RMS Mirnov coil signal, which is a measure of $\langle\dot B_{\theta}\rangle_{\text{RMS}}$. A point in (b) which clearly lies away from the main trend of the data is indicated by a red $\opentriangle$. Fits to the data in (b) both including and excluding this outlying data point are shown. Weighted correlation coefficients with and without this point are $\rho_{\text{inc}}=0.307$ and $\rho_{\text{exc}}=0.940$. Although a linear fit to the data is shown in (a), there is essentially no correlation ($\rho<0.1$). Shots which contribute to this figure are \#26857, \#26859 and \#26863, which are nominally identical to each other.}
\label{fig:correlation}
\end{figure}

Sufficient data have now been gathered with the FIDA spectrometer to allow studies of the ensemble of fishbones as well as individual events. Figure \ref{fig:deltabcorr} shows no correlation between the relative drop in core FIDA signal and the peak amplitude of the magnetic perturbation on the outboard midplane. Note that, since the fishbones contributing to this figure occurred in the flat-top phase of nominally identical shots, the absolute rather than relative perturbation amplitude is given. The variation in equilibrium magnetic field strength between different events is only a few percent. Although studies on NSTX \cite{Fredrickson2006} and JET \cite{PerezvonThun2012} have found some indication of correlation between magnetic perturbation amplitude and relative change in DD neutron rate, the correlations observed in those cases were very weak. In the data presented here, a much stronger correlation is observed between relative drop in FIDA signal and RMS amplitude of the time derivative of the magnetic field perturbation, as seen in Figure \ref{fig:bdotcorr}. These data are for a set of nominally identical shots; MAST is capable of producing highly repeatable plasmas, even down to the timing of individual fishbones being the same to within a few milliseconds. Note the clear outlier at $\langle\partial B/\partial t\rangle_{\text{RMS}}=46\usk\tesla\second^{-1}$, which is not associated with any indication of anomalous behaviour in the associated spectrogram. Including this data point in the fit, the weighted product-moment correlation coefficient supports correlation with only $64.1\%$ confidence. Excluding the point however, correlation is determined with at least $99.9\%$ confidence. The anomaly, which is derived from shot \#26859, may stem from the fact that a different passive view was used during this shot. For the other shots in Figure \ref{fig:correlation}, a passive view was used which looks just below the neutral beam, at the same toroidal location as the active view. For shot \#26859 however, a toroidally-displaced passive view was used instead. Toroidal asymmetries in the distribution of edge neutrals therefore affect the background subtraction of data from this shot.

Introducing data from other shots with different operating parameters almost completely removes the correlation observed in Figure \ref{fig:bdotcorr}. Three additional shots were included in the analysis: shot \#26789 is the higher density and higher current shot from which data are shown in Figures \ref{fig:trace26789} and \ref{fig:spectra26789}; shot \#27672 has one beam which starts later than the other, and suffers an internal reconnection event during 1-beam operation which causes a rapid drop in density and a reduction in normalised beta; and shot \#26855 is a 1-beam version of the shots based on \#26857. The unweighted correlation coefficient after adding fishbones from these shots to the dataset is 0.140.

\begin{figure}[p]
\begin{centering}
\includegraphics[width=0.6\textwidth]{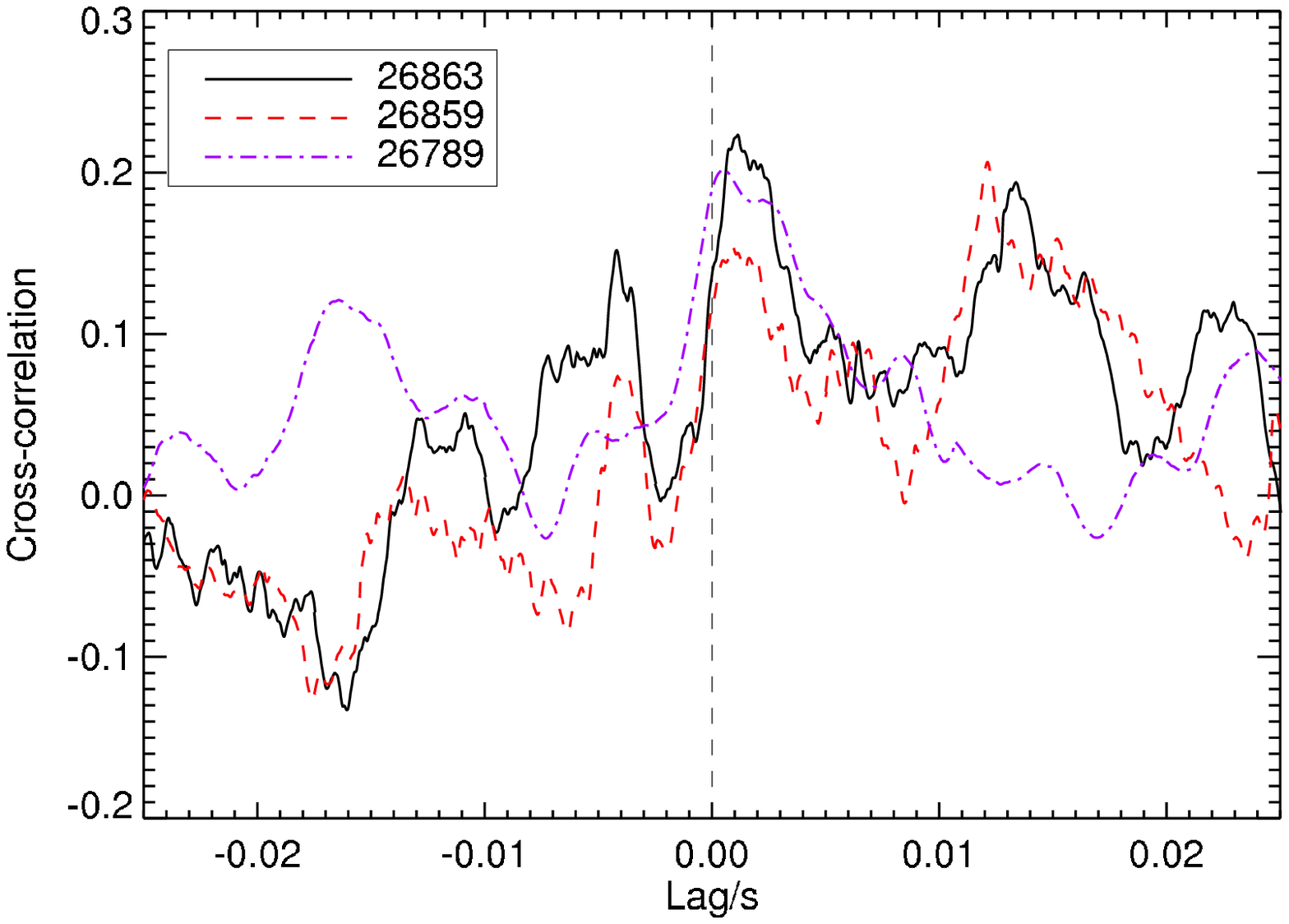}
\caption{Cross-correlation of fission chamber data with RMS amplitude of the outboard midplane Mirnov coil signal for time windows during which fishbones are active. Of the three MAST shots, \#26863 and \#26859 are nearly identical, whereas shot \#26789 exhibits higher density, normalised beta and plasma current.}
\label{fig:fc}
\end{centering}
\end{figure}

Data from the MAST fission chamber were also examined for these correlations, and although drops in the signal associated with fishbone bursts were observed, a methodical study of the type carried out with the FIDA data was deemed inappropriate because of the variable lag between the amplitude of the Mirnov coil signal and the drop in the neutron signal. It is however worth examining the fission chamber data for cross-correlation with the time derivative of the magnetic perturbation amplitude over the entire time window for which fishbones are active during selected shots. Cross-correlation between two data sets $n$ and $m$, each of $N$ elements, is defined as
\begin{equation}\label{eq:corr}
(n*m)(\tau) = \frac{\sum_{t=0}^{N-\tau-1}[(n_{t+\tau}-\bar n)(m_{t}-\bar m)]}
{\sqrt{\Bigl(\sum_{t=0}^{N-1}(n_t - \bar n)^2\Bigr)\Bigl(\sum_{t=0}^{N-1}(m_t - \bar m)^2\Bigr)}}
\end{equation}
where $n$ and $m$ in this case are the volume-integrated neutron rate and the RMS amplitude of the outboard midplane Mirnov coil signal and $\tau$ is the lag applied between the two signals. Figure \ref{fig:fc} shows cross-correlation as a function of lag for three shots. An apparently periodic variation in cross-correlation is indicative of the quasi-periodic occurrence of fishbones during the selected time windows. It is apparent in two cases that the largest correlation is obtained when the two signals are misaligned in time by a small lag of approximately $1\usk\milli\second$. Shot \#26859 however exhibits the largest correlation when the signals are offset by one whole fishbone `period' (approximately $12\usk\milli\second$), although the difference between the cross-correlation at $12\usk\milli\second$ lag and that at $1\usk\milli\second$ lag is very small. It is likely that the counterintuitive result in this shot stems from the presence of the outlier, seen in Figure \ref{fig:correlation}, which was observed to cause a small drop in FIDA signal relative to its magnetic perturbation amplitude. Although the values of cross-correlation in Figure \ref{fig:fc} are all $\ll 1$, the slow, quasi-periodic variation in cross-correlation is much larger than the small-scale point-to-point variation, indicating that this modulation in correlation is physically significant. It is unsurprising that the magnitude of the correlations themselves are small, given that the two signals are of a very different nature; the neutron rate tends to undergo positive and negative fluctuations about a slowly varying mean, whereas the RMS amplitude of the Mirnov coil signal has a flat, near-zero baseline where any variation must by definition increase the signal.

\begin{figure}[p]
\begin{centering}
\includegraphics[width=0.9\textwidth]{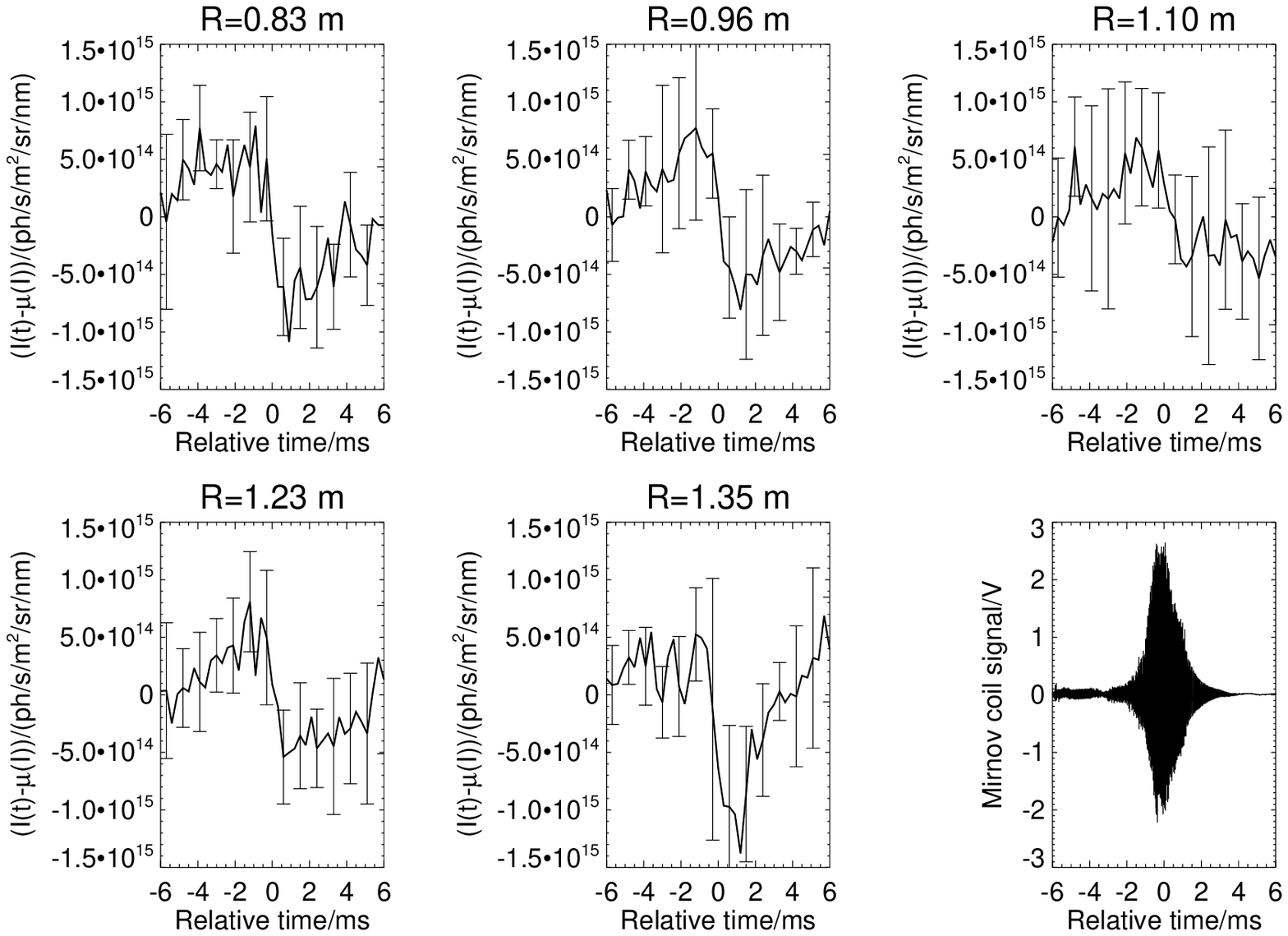}
\caption{FIDA signal intensity $I(t)$, $\lambda=660.7\usk\nano\metre$, at selected radii around the time of a `composite' fishbone event made up of the ensemble average of five events from nominally identical shots \#27919, \#27920 and \#27926. The mean of the data at each event was subtracted before averaging, but no further normalisation was performed. Error bars, plotted on every third data point, indicate the natural variation within the sample and are somewhat larger than those arising due to photon statistics and the systematic error of background subtraction, as seen in Figures \ref{fig:spectra26789} and \ref{fig:spectra26863}. The Mirnov coil trace of a typical fishbone from shot \#27919 is shown in the bottom-right panel. The magnetic axis for each of these shots, at the time of the fishbones selected for analysis, was located between $R=0.94\usk\metre \text{ and } R=0.96\usk\metre$.}
\label{fig:multiplot}
\end{centering}
\end{figure}

The final result presented here exploits the radial coverage of the FIDA diagnostic. Spanning from the tangency radius of the neutral beam at $R=0.77\usk\metre$ to the plasma edge at $R=1.41\usk\metre$, data across the entire outboard and part of the inboard midplane may be acquired in a single shot. The effect of fishbones on the FIDA emissivity at selected radii is presented in Figure \ref{fig:multiplot}. This plot incorporates FIDA data from a number of fishbones which occurred during several nominally identical shots. The resulting `composite' event gives us an idea of the typical behaviour and range of variation in this behaviour due to individual fishbones. It is observed that large drops in the magnitude of the signal occur across the plasma radius both inside and outside the magnetic axis, situated in this case near $R=0.96\usk\metre$. This is consistent with the expected kink-ballooning mode structure illustrated in Figure \ref{fig:structure}, which has a large amplitude as far out as the $q=2$ surface, located in this case at around $R=1.30\usk\metre$ on the outboard midplane. It is interesting to note that the drop in signal is weaker at $R=1.10\usk\metre$ than at either of the adjacent radii. For these shots, at the times of the selected fishbones, $R=1.10\usk\metre$ corresponds approximately to the outboard radial location of $q_{\min}$. In all cases contributing to this figure, $q_{\min}>1$. The data at $R=1.35\usk\metre$ are once again affected by incorrect subtraction of passive FIDA from edge neutrals; the rapid burst of passive FIDA causes a transient drop in the net signal in this case. A toroidally-displaced passive lens was used for these shots, which detects a larger passive FIDA component at the relevant wavelength than the active view, probably due to toroidal asymmetries in the distribution of neutrals at the plasma edge. Note that this effect was also seen in data from shot \#26859, shown in Figure \ref{fig:correlation}. This contrasts with the behaviour of the signal in Figure \ref{fig:trace26789} and Figure \ref{fig:trace26863} in which the spikes in passive signal at the edge are clearly larger in the active than in the passive view.

\section{Conclusions}\label{sec:conc}
Transient drops in core FIDA signal have been observed associated with low-frequency fishbone-type MHD modes in MAST. Each drop, associated with a chirping mode, is followed by a recovery phase where the emissivity grows to an apparently critical value before the next fishbone is triggered. The modes themselves are observed even in the absence of a $q=1$ surface inside the plasma, and the drops in FIDA signal extend from the magnetic axis out to at least mid radii. These observations are consistent with the $n=1$ infernal kink-ballooning mode structure predicted theoretically \cite{Pinches2012}. The relative amplitude of the drop in FIDA signal is observed to be correlated with the RMS amplitude of the time derivative of the magnetic perturbation under certain scenarios, but this correlation is lost when fishbones under a range of operating scenarios are considered. The behaviour of the FIDA signal is seen to be consistent across fishbones at different times within the same shot, as well as across nominally identical shots. During periods of quasi-periodic fishbone activity, the volume-integrated neutron rate also shows a quasi-periodic variation in correlation with the RMS magnetic coil signal. Further developments in the modelling of passive FIDA emission from edge neutrals are required before accurate information on fast-ion redistribution can be extracted from FIDA emission near the plasma edge, but preliminary results are consistent with those from core channels. In particular, fishbone-induced redistribution is seen in both core and edge channels to extend over at least the range of wavelengths corresponding to $E_{\min}=30\usk\kilo\electronvolt$ up to $95\%$ of the full injection energy, regardless of whether the two beams have the same or different injection energies.

Although MAST is scheduled to undergo a major upgrade between late 2013 and early 2015, one further experimental campaign remains before the shutdown. Coordinated experiments with the FIDA diagnostic and collimated neutron camera are planned, and should extend our understanding of the effects of MHD instabilities, in particular energetic particle modes, on fast ions in spherical tokamaks. It is hoped that theoretical and numerical modelling will continue to progress to allow a deeper understanding of the effect of repeated chirping modes of the type reported here. Such modelling has been identified as an outstanding challenge in the field \cite{Breizman2011}. The need to identify the particular portion of phase space which resonates with, and hence drives, the fishbones in accordance with the theory outlined in Reference \cite{Kolesnichenko2000} remains an outstanding problem. The present work also poses a challenge to explain the correlation between changes in fast-ion density and the peak \emph{time derivative} of the magnetic perturbation, rather than the perturbation amplitude itself. In addition to studying the redistribution arising from these modes, it is proposed to empirically identify regimes with high plasma performance in which this fast-particle MHD is suppressed. It then remains to be seen if, under such scenarios, the rate of fast-ion redistribution conforms completely to neoclassical predictions.

\ack The assistance of S. Sharapov in interpreting the results of MISHKA modelling is gratefully acknowledged. This work was funded partly by the RCUK Energy Programme under grant EP/I501045 and the European Communities under the contract of Association between EURATOM and CCFE. The views and opinions expressed herein do not necessarily reflect those of the European Commission or of the ITER Organization.

\clearpage
\section*{References}
\bibliographystyle{unsrt}
\bibliography{library}

\end{document}